\documentclass{article}
\usepackage{spconf,amsmath,graphicx,hyperref,color}

\title{Evaluating pretrained speech embedding systems for dysarthria detection across heterogenous datasets}

\name{{\em Lovisa Wihlborg$^1$, Jemima Goodall$^1$, David Wheatley$^1$, Jacob J. Webber$^1$, Johnny Tam$^{2,4}$,} \\
{\em Christine Weaver$^{2,4}$, Suvankar Pal$^{2,4,5}$, Siddharthan Chandran$^{2,4,5}$,} \\
{\em Sohan Seth$^{3}$, Oliver Watts$^{1,2}$, Cassia Valentini-Botinhao$^{1}$}}
\address{$^1$SpeakUnique Ltd., UK, $^2$Anne Rowling Regenerative Neurology Clinic, \\ University of Edinburgh (UoE), UK, $^3$Institute of Adaptive and Neural Computation, UoE, UK, \\ $^4$Euan MacDonald Centre for MND Research, UoE, UK, $^5$UK Dementia Research Institute, UK}

\begin{document}
\ninept
\maketitle
\begin{abstract}
We present a comprehensive evaluation of pretrained speech embedding systems for the detection of dysarthric speech using existing accessible data.
Dysarthric speech datasets are often small and can suffer from recording biases as well as data imbalance.
To address these we selected a range of datasets covering related conditions and adopt the use of several cross-validations runs to estimate the chance level. 
To certify that results are above chance, we compare the distribution of scores across these runs against the distribution of scores of a carefully crafted null hypothesis.
In this manner, we evaluate 17 publicly available speech embedding systems across 6 different datasets, reporting the cross-validation performance on each. 
We also report cross-dataset results derived when training with one particular dataset and testing with another. 
We observed that within-dataset results vary considerably depending on the dataset, regardless of the embedding used, raising questions about which datasets should be used for benchmarking.
We found that cross-dataset accuracy is, as expected, lower than within-dataset, highlighting challenges in the generalization of the systems.
These findings have important implications for the clinical validity of systems trained and tested on the same dataset. 
\end{abstract}

\section{Introduction}

Dysarthria is a motor speech disorder characterised by disordered articulation (slurred speech) and caused by damage to the nervous system, which leads to paralysis, slowing, or incoordination of the muscles of articulation. Causes include neurological disorders such as stroke,  amyotrophic lateral sclerosis (ALS), Parkinson's disease (PD) and developmental disorders such as cerebral palsy (CP). In particular, neurodegenerative disorders are a significant and increasing health burden globally, with urgent and unmet needs for accessible and accurate diagnostic and monitoring methods \cite{ferrari2024global}. Speech has great potential to provide a non-invasive, remotely collected, and objective digital biomarker in these conditions. Speech processing and machine learning methods have recently shown great promise in both disease detection and severity prediction from only the speech signal \cite{bowden2023systematic, de2020artificial}. However, clinical translation and adoption of these methods will depend on robust evaluation of speech processing methods and clinical validation. 
% \blfootnote{Correspondence email: lovisa@speakunique.co.uk}

% Problem: how to reliabily evaluate speech embeddings 
% There has been much recent interest in using speech to monitor and characterise speaker health, and many systems and approaches have been proposed \cite{bowden2023systematic, de2020artificial}.
The success of speech embedding systems trained in a generic way when applied to specific tasks has led to an interest in applying them also to speaker health evaluation \cite{javanmardi2024pretrained}.
These systems are relatively easy to engineer as they rely on publicly available tools that often do not require in-depth expertise of either the task or the data.
However, it is  challenging to reliably evaluate their performance on health monitoring tasks. Results in this domain are often reported on closed test sets that cannot be shared for ethical and legal reasons. Where results are published using accessible data, their reliability is compromised by the small size and known artefacts of the available open datasets. Moreover, results are often presented without sufficient information about how data is preprocessed and partitioned, meaning that even when data can be accessed, it is doubtful whether results can be replicated. 

% Previous work - more dataset, challenges, dataset evaluation
Several lines of work seek to remedy this situation. 
The creation of new larger and better open datasets (such as \cite{speech_access_project})
%(ADDI 
is welcome, as are shared tasks that define firmer train-set partitions for existing datasets (e.g.\ \cite{luz21_interspeech}).
%However, it is not clear how different datasets can be combined or if they should be combined at all.
A separate line of work validates specific existing datasets and sheds light on the reliability of results obtained from them.
Datasets can be flawed in many different ways, and two particular issues are to do with data balance and coverage. Data must be balanced in such a way that a system is  incapable of learning to discriminate between underlying classes using factors that are not relevant to the task in a generalisable way. Balance is achieved by ensuring similar data acquisition procedures across classes, both speaker-intrinsically in terms of the demographics of the participants (age, gender, accent) and also extrinsically -- in the case of audio data, for example, ensuring that no consistent differences of recording setup (microphone, environment) correlate with the classes or health measures to be detected.

% Our angle -- more dasets, multiple partitions, null distribution, non-speech
In line with these considerations we present a large and comprehensive evaluation of systems for dysarthria detection. 
Rather than relying on one or two potentially problematic datasets, we use multiple datasets covering related conditions. 
Furthermore we used repeated $k$-fold cross-validation, varying fold splits across runs. To verify that results are significantly above chance level we contrast the distribution of scores across these runs to a null hypothesis distribution of each feature--task combination.
We evaluate 17 speech embedding systems across six different datasets and present cross-validation within-dataset performance of each system on each dataset.
We also present cross-dataset results using a selection of closely related datasets. These results are obtained by training systems on one complete dataset and testing on another.

\begingroup
%%% Decreasing space betweeen cols.
\setlength{\tabcolsep}{5pt} % Default value: 6pt
\begin{table*}[t]
\centering
\caption{Datasets used in evaluation. Selected conditions: healthy control (HC), cerebral palsy (CP), Parkinson’s disease (PD), Huntingon’s disease (HD), amyotrophic lateral sclerosis (ALS), $^*$refers to: peripheral neuropathy, myopathic or myasthenic lesions. Access: academic non-profit purposes (A), commercial and non-commercial (C+NC). \#Speakers after balancing.}
\begin{tabular}{l l l l l l l}
\hline
Dataset	                   & Selected conditions & Language       & \#Speakers (HC) & Elicitation tasks & Access & Preparation  \\ \hline
EWA	\cite{EWA24}           & HC/PD	    & Slovak         & 190 (95)       & Mixed           & free license (C+NC) & Balanced for class \\
EasyCall \cite{EasyCall21} & HC/PD/HD/ALS/$^*$ & Italian    & 45 (24)          & Commands        & CC BY-NC            & Balanced for class  \\
Neurovoz \cite{NeuroVoz24} & HC/PD      & Spanish        & 108 (55)        & Mixed           & CC BY-NC-ND 4.0     & -   \\
SSNCE \cite{ssnce}         & HC/CP      & Tamil	         & 20 (10)         & Mixed           & free license (A)    & Balanced for class    \\
TORGO \cite{Torgo12}	   & HC/CP/ALS  & N.A. English   & 15 (7)        & Mixed           & free license (A)    &  -    \\ 
UASpeech \cite{UASpeech08} & HC/CP	    & N.A. English	 & 28 (13)     & Word repetition & CC BY 4.0           & -  \\
\hline
\end{tabular}
\label{table:datasets}
\end{table*}
\vspace{-10pt}
\endgroup

\section{Background}

\subsection{Datasets of dysarthric speech} \label{Background:subsect-datasets} % Something more general about datasets
Many datasets containing recordings of dysarthric speech have been reported, and several can be obtained relatively straightforwardly for research \cite{speech_access_project, EWA24,EasyCall21,NeuroVoz24, ssnce, Torgo12, UASpeech08}. They differ in terms of conditions (what caused the impairment), language, number of speakers, amount of data collected per speaker, type of elicitation tasks and access. Motivations for creating and releasing such datasets include, among other things, the improvement of automatic speech recognition (ASR) of dysarthric speech \cite{speech_access_project, Torgo12} and automatic diagnosis \cite{EWA24, UASpeech08}.

% Biases and how to check them
Speech datasets can be imbalanced in terms of speaker characteristics (i.e.\ age, gender, accent and health condition) and speaker-extrinsic factors such as recording conditions (microphone, environment, data format). 
To ensure that a system discriminates between underlying classes using only task-relevant factors that are generalisable, it is crucial to understand and address these imbalances where possible.
While some of the biases can be addressed straightforwardly (e.g.\ by choosing a subset of the dataset with equal numbers of males and females) others are harder to revert without loss of performance (e.g.\ removing noise and reverberation). The authors in \cite{schu2023using} found that embedding-based classification systems were able to diagnose speakers successfully based just on the `silent' (i.e.\ non-speech) portions of the TORGO \cite{Torgo12} and UASpeech \cite{UASpeech08} datasets, due to systematic differences in recording conditions between the healthy and impaired groups. Authors of \cite{liu2024clever} found a similar effect when classifying the `silent' regions of the `picture description task' recordings of the Pitt Corpus from DementiaBank \cite{liu2024clever}.

\subsection{Speech embeddings}

Speech embedding systems generate a continuous fixed or variable-length representation that characterises the speech signal. 
While traditional speech descriptors have been based on signal processing modules such as pitch and spectral envelope extractors \cite{disvoice,egemapsv2,vadfeat}, speech embedding systems tend to refer to systems based on data-driven statistical models, nowadays mostly neural networks \cite{wav2vec2bert,wav2vec2conformer,unispeech,unispeechPaper,wav2vec2,speechbrainenc-decasr,vggish,TitaNet,speechbrain,ecapa,tddxvector,wespeaker,efficienttdnn,resemblyzer}.
The most common architecture of a speech embedding system consists of an encoder that converts the speech waveform into a compact representation, and optionally a network or other mechanism that summarises it over time into a fixed-length representation.
Speech embedding systems tend to be trained with thousands of hours of speech from several speakers and in some cases, multiple languages. 
% Add citations to following sentences
While speech descriptors derived from signal processing extract features that are to some extent interpretable, speech embedding systems are not usually designed with interpretability in mind.
These systems can be trained to discover representations that aim to characterise speech in a task-agnostic manner relying solely on the speech waveform or in a task-dependent manner by learning from additional information sources such as speaker identity and text.
Depending on their pretraining task we can categorise these systems as self-supervised, based on speaker verification (SV), or ASR tasks.

\begingroup
%%% Decreasing space betweeen cols.
\setlength{\tabcolsep}{5pt} % Default value: 6pt
\begin{table*}[t]
\centering

\caption{Systems used for evaluation grouped by pretraining task (self-supervised, ASR, SV, none) and ordered by number of parameters (Params.). Architecture: convolutional neural network (CNN), transformer (Tr), time delay neural network (TDNN), statistics pooling (Sp), attention pooling (Ap). Pretraining task: connectionist temporal
classification (CTC), additive angular margin (AAM), angular margin softmax (AMS), generalized end-to-end (GE2E). Data: BABEL (B), CommonVoice (CV), GigaSpeech (GS), Librispeech (Ls), Libri-Light (LL), VoxPopuli (VP), VoxCeleb1 (VC1), VoxCeleb2 (VC2), YouTube-8M (YT). Size: $^*$indicates averaged across time.}

% Spreadsheet of systems with links:
% https://speakunique-my.sharepoint.com/:x:/g/personal/cassia_speakunique_co_uk/EamVOauUGRpOkt6E3crzwHsB_pl4Qn1YCP-Rh1xuVynSVQ?e=LbNhnB

\begin{tabular}{l l l l l l l l l }
\hline
System	& Architecture & Params. (M) & Pretraining task & Hours (Data) & Speakers  & Langs. & Size \\ \hline

%% Self-supervised
Wav2Vec2Bert \cite{wav2vec2bert}& CNN+Tr+x-vector & 635 & Contrastive & 4.5M & n/a & 143 & 512 \\
Wav2Vec+Conf \cite{wav2vec2conformer} & Conformer & 618 & Contrastive & 960 (Ls) & 2,484  & EN & 512 \\
UniSpeech-SAT \cite{unispeechPaper} & CNN+Tr & 94.68 & Contrastive & 94k (LL,GS,VP,VC1) & $>$9,963 & EN & 512 \\ % (7,439+?+1,313+1,211)
 \hline

%% ASR
UniSpeech \cite{unispeech} & CNN+Tr & 300 & Contrastive+CTC & 1.5k (CV,TIMIT) & 60k & EN & 1024$^*$ \\
Wav2Vec2-XLSR \cite{wav2vec2}   & CNN+Tr & 300 & Contrastive+CTC & 56k (MLS+CV+B) & n/a & 53 & 512$^*$ \\
CRDNN+CTC \cite{speechbrainenc-decasr} & CNN+LSTM & 172 & CTC & 960 (Ls) & 2,484 & EN & 512\\  \hline

% Audio classification
VGGISH \cite{vggish}       & CNN & 133 & Cross-entropy & $<$350k (YT subset) & n/a & n/a & 128$^*$\\

%% SI systems
TitaNet \cite{TitaNet}	        & CNN+Ap & 23 & AAM & 3k (Ls,VC12,others) & 16.6k  & EN & 192 \\ % (VC1, VC2, Fisher, Switchboard, Librispeech, SRE) % additive angular margin loss
ResNetTDNN \cite{speechbrain}   & ResNet TDNN+Sp & 22.1 & AMS & 7,794 (VC1,VC2)  & 7,205 & EN & 256 \\ %  Additive Margin Softmax
ECAPA-TDNN \cite{ecapa}         & TDNN+Ap & 14.7 & AMS &  7,794 (VC1,VC2)  & 7,205 & EN & 192 \\ %  Additive Margin Softmax
x-vector \cite{tddxvector}       & TDNN+Sp & 8.2 & Cross-entropy & 7,794 (VC1,VC2) & 7,205 & EN & 512 \\ % Categorical Cross-Entropy Loss.
WeSpeaker \cite{wespeaker}      & ResNet34+Sp & 6.6 & AAM & 2,442 (VC2) & 5,994 & EN & 256 \\ % AAM 
EfficientTDNN \cite{efficienttdnn} & TDNN+Ap & 5.79 & AAM & 2,442 (VC2) & 5,994 & EN & 192 \\ % AAM
Resemblyzer \cite{resemblyzer}  & LSTM & 1.4 & GE2E & 960 (Ls) & 2,484 & EN & 256 \\ % GE2E

% additive angular margin (AAM), generalized end-to-end (GE2E)

\hline
%% DSP
DisVoiceProsody \cite{disvoice} & - & - & -& - & - & - & 103 \\
eGeMAPSv2 \cite{egemapsv2} & - & - & - & - & - & - & 88 \\
DigiPsychProsody \cite{vadfeat} & - & - & - & - & - & - & 21  \\
\hline
\end{tabular}
\label{table:systems}
\end{table*}
\endgroup

\begin{figure*}
    \centering
    \includegraphics[width=1.0\linewidth]{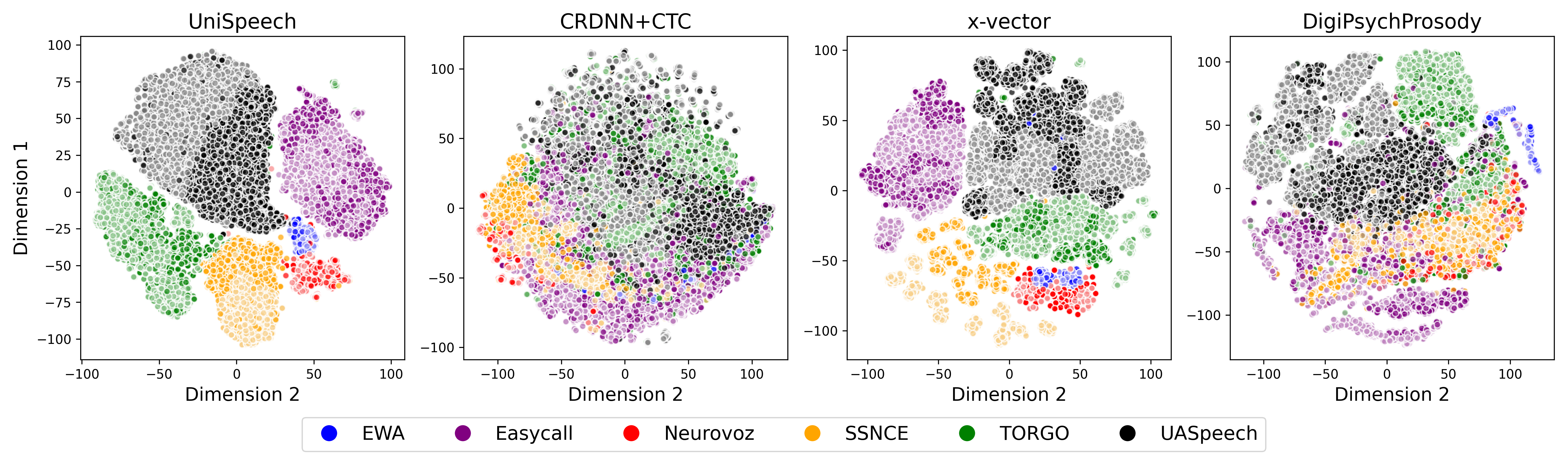} % this png was generated on Preview from the pdf using 100 pixels/cm as resolution
    \vspace{-15pt}
    \caption{t-SNE visualisation of speech embeddings for healthy control (light shade) and dysarthric (dark shade) speakers.}
    \label{fig:t-sne}
    \vspace{-10pt}
\end{figure*}

\begin{figure*}[t]
\vspace{-20pt}
\centering
\includegraphics[clip, width=0.49\textwidth]{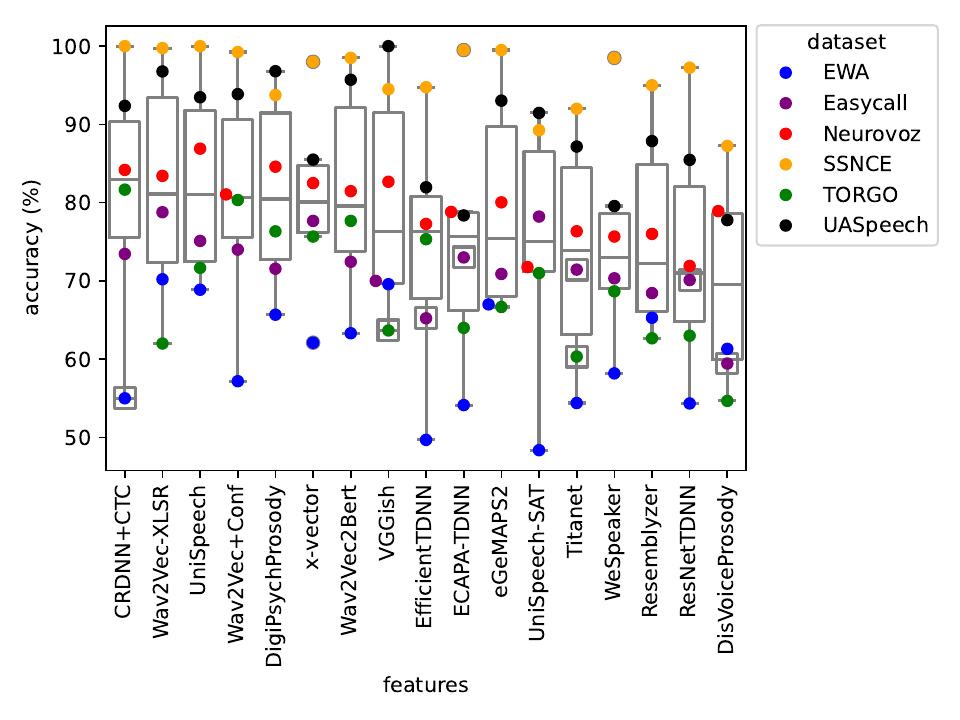}
\includegraphics[clip, width=0.49\textwidth]{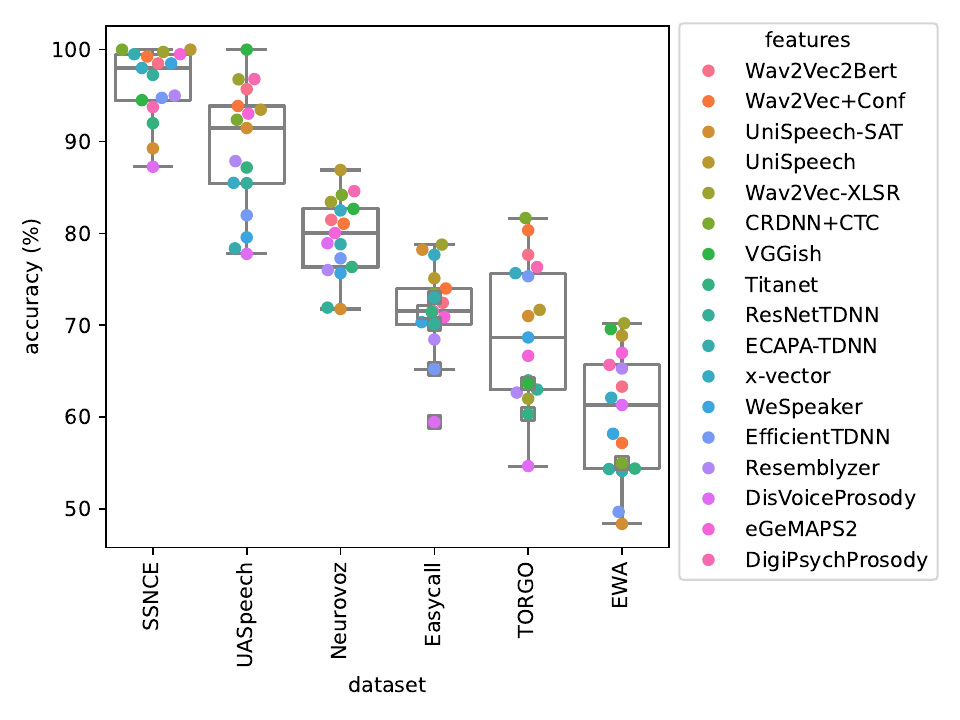}
\vspace{-15pt}
\caption{Within-dataset accuracy broken down by feature (left) and dataset (right).} \label{fig:within_dataset}
\vspace{-10pt}
\end{figure*}

\section{Evaluation}

\subsection{Data, tasks and systems}\label{section:data}
% Data
Table \ref{table:datasets} presents a summary of the datasets we used. 
The primary reason for this selection was to utilise publicly available datasets containing dysarthric speech caused by neurological conditions. As a reflection of that, the range of medical conditions covered by each of these datasets differs. It is worth noting that the chosen datasets cover a range of languages, potentially enabling the evaluation of language-independent systems.
% Tasks
We defined a binary classification task for each dataset (discriminating healthy from dysarthric speech). When multiple dysarthria-associated conditions existed in a particular dataset, all of them were grouped as part of the speech dysarthric class. For all datasets we ensured similar-sized groups for the 2 classes; the table notes where we needed to select a subset of speakers to create this balance. We made use of sex and age information where available to ensure comparable distribution in both groups.

% \input{datasets} %%% Short description of each

% Systems
Table \ref{table:systems} shows the details of the systems we chose to evaluate. 
These systems were selected as publicly available exemplars of a range of successful techniques.
These include neural speech embedding systems trained on different tasks (self-supervised, ASR and speaker verification) as well as systems based on signal processing (bottom section of the table).
Fig. \ref{fig:t-sne} presents t-SNE visualisations of four notable speech embedding systems.
Datasets are clearly distinguished in the UniSpeech and x-vector spaces, and somewhat in DigiPsychProsody.
The x-vector space reveals smaller clusters that closer inspection shows correspond to different speakers.
% ^---- TODO: need to check this!
In contrast, CRDNN+CTC space shows less distinct dataset separation and no speaker clusters, although the healthy and dysarthric separation seems less clear.
% The healthy and dysarthric conditions seem better separated in the UniSpeech space, particularly for the UASpeech and the SSNCE datasets.

\subsection{Feature extraction}

Each system in Table \ref{table:systems} is employed for feature extraction, generating 17 different feature sets. % making the connection between the words systems and features
Prior to feature extraction all data was resampled to $16$\,kHz where necessary.
Representations were then extracted from each recording using the code and pretrained models from the publicly available implementations (see links in references). For representations consisting of a single embedding whose fixed dimension is invariant to audio input length (dimensions are presented in the last column of Table \ref{table:systems}), these features were then used directly to represent the recording. For representations with a time axis, we reduced each to a single vector by taking the mean over time \cite{javanmardi2024pretrained}. Informal experimentation suggested that a different choice of arithmetic summary (median, etc.) had little impact on the results obtained.
When a single vector representation had been produced for each recording, they were used along with the binary labels indicating healthy and dysarthric status to train classifiers. 

\subsection{Cross-validation and chance level estimation}

%To evaluate the usefulness of each of the representations on the tasks we defined over the various datasets we adopt the following procedure for building a classifier on each feature--task pair.

For our machine learning approach we chose to use Random Forest \cite{Breiman01} throughout, as this method performs solidly across varied datasets without requiring extensive hyperparameter tuning. 
%[Description of actual forest building? complete splitting; m-value is sqrt of n features; gini-criterion]. 
We used the RandomForestClassifier implementation in scikit learn with its default configuration, except that the number of trees per forest was set to 1000 and fixed random seeds were used for replicability.
% Cross-validation
Training and evaluation was carried out everywhere with 20 runs of 5-fold cross-validation.
The cross-validation was performed by grouping a speaker's samples so that -- in datasets where a speaker has multiple recordings -- there was never a single speaker present in both train and test splits. Stratification was used so that at least one example of each class was present in each test split.
In datasets where a speaker has multiple recordings, a speaker's predicted label was selected by majority voting over the predicted label for each recording.
% Chance level estimation
The way all tasks are set up to be balanced with regard to classes means that the chance level for most metrics reported will be around $0.5$. To account for slight variations and also to generate a chance distribution for each feature--task combination, we performed a permutation analysis (see e.g. \cite{yu2003resampling}).
Speakers' labels were shuffled, respecting the constraint that all examples from a speaker should share the same label. Then the same 20 runs of 5-fold cross validation were performed using identical data but with the shuffled labels. The resulting 20 points are treated as a sample from the chance (null) distribution. We then used a Welch’s one-tailed t-test to determine whether the average score for a feature was significantly higher than the average chance level score, given the distribution of interest and its null distribution.

\subsection{Within-data results}

For conciseness, we consider only accuracy in our analysis; this is possible because we ensured class balance in the datasets used (Section \ref{section:data}). 
Fig. \ref{fig:within_dataset} presents accuracy results organised per system (left) and dataset (right). 
Results are shown via boxplots of the distribution of the mean accuracy scores as well as dots indicating these mean accuracy score obtained in each of the 6 datasets (left) and by each of the 17 systems (right). 
Boxes show the quartiles of the 6 point distribution (left) and the 17 point distribution (right).
All except 8 of the 102 system--dataset accuracy scores were found to be significantly higher than chance (with $\alpha$ set to $0.05$, and Bonferroni correction for multiple comparisons); this is indicated by a square shape in Fig.\ \ref{fig:within_dataset}. The systems and datasets are ordered based on the median score, represented by the horizontal line in the boxplots. The legend and colours of the systems in the right hand side plot of Fig. \ref{fig:within_dataset} are ordered according to the groups in Table \ref{table:systems}.

% Systems performance
We can see that speech embedding systems trained with an ASR task performed the best on average. The x-vector presented the lowest variability in performance across the different datasets considering the interquartile range. DigiPsychProsody performance is close to the best neural systems despite using a much smaller feature set and not depending on hundreds of hours of data.
% Performance across datasets
Accuracy results vary drastically depending on the datasets. All systems perform very well on the SSNCE dataset with most accuracies above 95\% and substantially worse on EWA where accuracies are mostly below 65\%. The systems' performances seem more consistent on the Easycall dataset even though it contains more health conditions than other datasets. Performance varies the most on the TORGO dataset, potentially due to the small number of speakers and the known recording condition biases.

\newif\ifsilenceresults
\silenceresultsfalse % swtich this between true/false to include silence results
\ifsilenceresults
    \subsection{Confounding due to recording conditions}\label{section:silence}
    Previous work has highlighted the confounding influence of recording conditions in health-related speech datasets, including 2 of the datasets used in the current work (UASpeech and TORGO) \cite{schu2023using}. Both \cite{schu2023using} and \cite{liu2024clever} show that it is possible to obtain health classification accuracies significantly better than chance based only on non-speech segments of recordings. In such cases, accuracies found using the speech data are likely inflated, and systems will not be robust to small changes in data capture. We approximately replicate those results here, and extend them to 2 more datasets. 
    
    Additional versions of datasets were produced  containing only segments determined by the voice activity detector in RNNoise \cite{rnnoise} to be non-speech. A segment was classified as non-speech when it was longer than $50$\,ms and when the probability of voice activity given by the model was below $20$\% over all its frames.
    Table \ref{table:non-speech} presents the accuracy results we obtained for TORGO, UASpeech, SSNCE and NeuroVoz.
    \color{red}
    We were not able to detect sufficient non-speech segments from Easycall and EWA to run this analysis for those datasets.
    Performance using non-speech segments is especially high for SSNCE, TORGO and UASpeech. Notably, for TORGO, it exceeds the accuracy obtained using the full dataset.
    This might explain why systems consistently perform well on both SSNCE and UASpeech, but show considerable variability with the TORGO dataset.
    \color{black}
    
    \begin{table}[t]
    \centering
    \caption{Accuracy scores on speech and non-speech, averaged across all systems. Standard deviation in parenthesis. }\label{table:non-speech}
    \begin{tabular}{l l l}
    \hline
    Dataset                     & Full Data     & Silence only \\ \hline
    Neurovoz                    & 80.84 (6.64) & 70.41 (8.17) \\ \hline
    SSNCE	                      & 93.53 (9.79) & 77.9 (19.21) \\ \hline
    TORGO                       & 70.97 (12.18) & 82.64 (14.10) \\ 
    TORGO \cite{schu2023using}  & 68.57 (6.58)  & 79.68 (3.69) \\  \hline
    UASpeech                    & 88.86 (10.09) & 81.5 (23.52) \\
    UASpeech \cite{schu2023using} & 89.79 (3.74) & 90.63 (3.86)  \\ 
    \hline
    \end{tabular}
    \vspace{-10pt}
    \end{table}
\fi 

\begin{figure}[t]
\centering
\includegraphics[clip, width=0.23\textwidth]{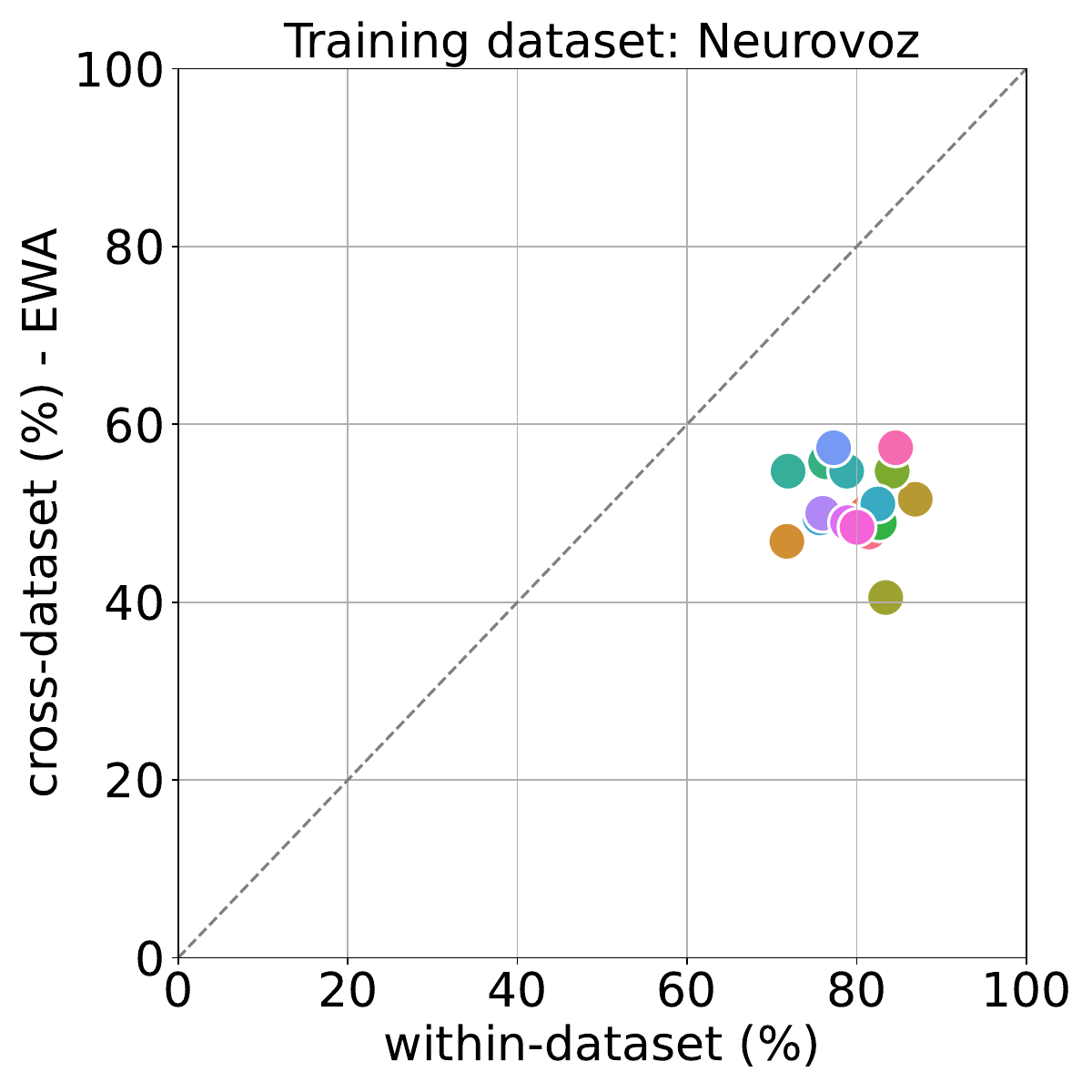}
\hspace{5pt}
\includegraphics[clip, width=0.23\textwidth]{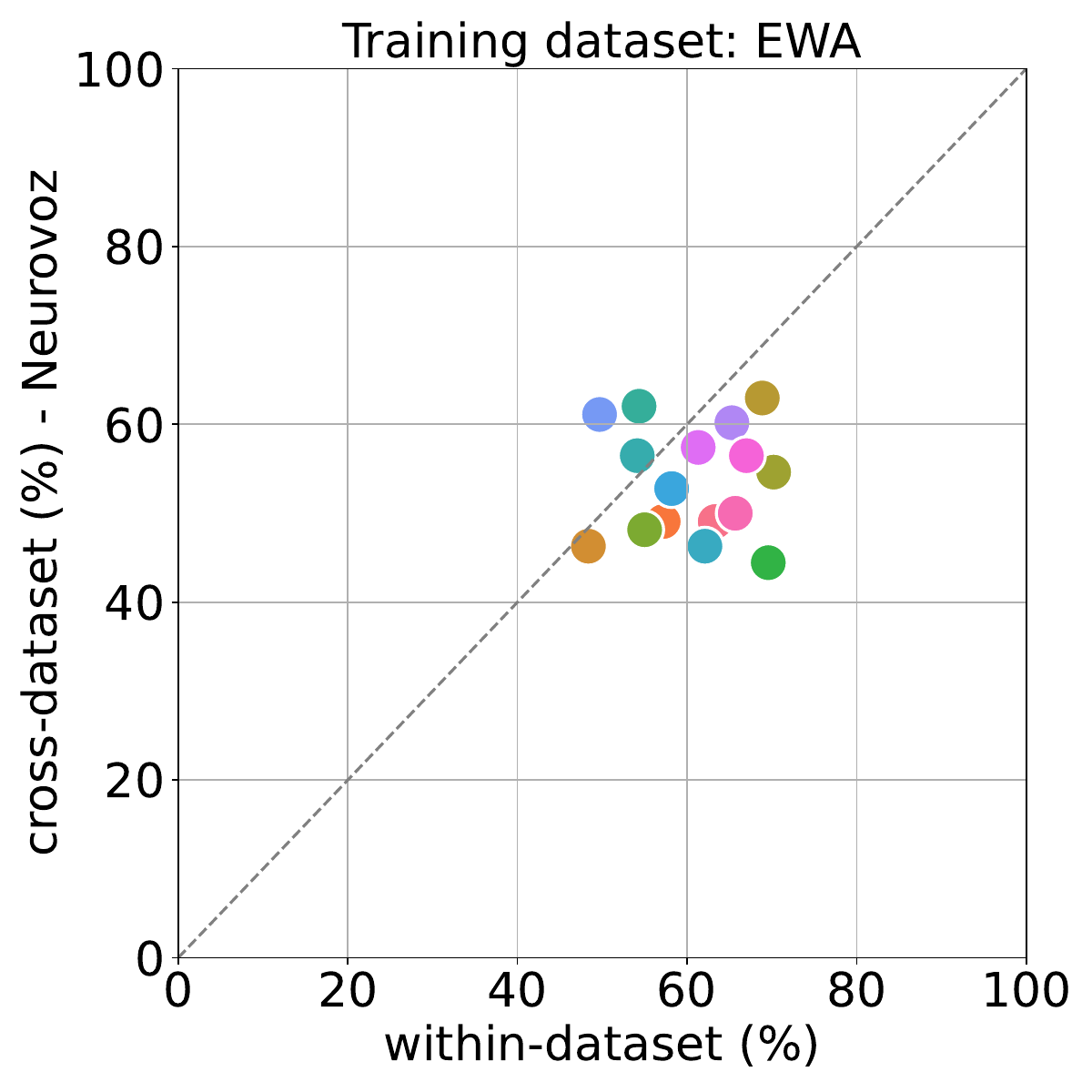}
\vspace{-10pt}
\caption{Cross-dataset versus within-dataset accuracy for Neurovoz (left) and EWA (right). System colours are the same as in Fig. 2.} \label{fig:lodo}
\vspace{-15pt}
\end{figure}

\subsection{Cross-dataset results}
Evaluation presented here thus far has been restricted to be within dataset: systems are trained and tested via cross-validation on samples from the same dataset. 
Aggregation of these scores across multiple datasets mitigates to some extent dataset biases when comparing systems. 
However, known issues with two datasets used (UASpeech and TORGO \cite{schu2023using}) which we have been able to replicate, may also be present in others, potentially inflating scores.
A crucial next step is to include assessment of system performance when tested on datasets unseen in training, which will enable conclusions that are robust to confounding due to data collection artefacts.
% -- both speaker-extrinsic variation of the sort revealed in Section \ref{section:silence} as well as speaker-intrinsic confouding such as imbalances of demographic attributes between classes. 
Such cross-datasets analysis might take the form of \textit{leave one dataset out} evaluation, where systems are trained on several pooled datasets and tested on a completely held-out dataset \cite{kubinski2022benchmark}. This is challenging in general due to the difficulty of successfully aggregating speech data from different sources \cite{lee2019learning}, and specifically with the publicly available speech datasets used here which are heterogenous in terms of the medical conditions they cover. However, we make a first step in this important direction by focusing on two datasets (EWA and Neurovoz) that both have a relatively large number of speakers and are homogenous in terms of the condition they cover (PD). 

Fig. \ref{fig:lodo} shows cross-dataset versus within-dataset accuracy of models trained on Neurovoz (left) and EWA (right). The within-dataset accuracy is the cross-validation accuracy reported in Fig.\ref{fig:within_dataset}.
Cross-dataset accuracy is derived by training a model on the entirety of one dataset and testing it on the other dataset.
It can be seen that accuracy drops when moving from within- to cross-dataset evaluation 
($79.62\% \xrightarrow{}51.08\%$ and $60.27\% \xrightarrow{} 54.08\%$  when training on Neurovoz and EWA, respectively).
% Neurovoz 79.62 (4.35) -> 51.08 (4.35)
% EWA 60.27 (6.96) -> 54.08 (6.29)

\section{Conclusions and future work}

% Brief summary of paper and findings
We presented a large-scale evaluation of 17 speech embedding systems on 6 different publicly available datasets for dysarthric speech detection. We reported accuracy results within-dataset, obtained via cross-validation, as well as cross-dataset results on a selection of the datasets.
We observed that, regardless of the classification system, certain datasets are consistently and drastically easier to classify than others. 
This could indicate that the task in these datasets is genuinely easier (e.g., speech is more dysarthric and distinct from healthy speech), or that there is a hidden bias in the dataset making the task easier. 
In either case, our results raise questions about how systems should be benchmarked and whether a single dataset is sufficient.
The cross-dataset results we report are, as expected in the settings we evaluate, worse than within-dataset results. This highlights the need for developing representations and models that are more robust to dataset specific confounding factors, and that are therefore more generalisable and appropriate for real life settings.

\vspace{5pt}

{\footnotesize%
\noindent
\textbf{Acknowledgement:} %
This work is supported by NEURii, a collaborative partnership involving the University of Edinburgh, Gates Ventures, Eisai, LifeArc and Health Data Research UK (HDR UK).}

\bibliographystyle{IEEEtran}
\bibliography{refs}

\end{document}